\documentclass[fleqn,10pt]{wlscirep}
\usepackage[utf8]{inputenc}
\usepackage[T1]{fontenc}
\usepackage{color}
\usepackage{soul}
\usepackage{comment}

\title{Understanding European Integration with Bipartite Networks of Comparative Advantage}

\author[1,2,$\ast$]{Riccardo Di Clemente}
\author[3,4,$\ddag$]{Balázs Lengyel}
\author[5]{Lars F. Andersson}
\author[6]{Rikard Eriksson}
\affil[1]{Exeter University, Department of Computer Science, Exeter, EX4 EPY, United Kingdom.}
\affil[2]{The Alan Turing Institute, London NW12DB, United Kingdom.}
\affil[3]{Agglomeration and Social Networks Lend{\"u}let Research Group, Centre for Economic and Regional Studies, E{\"o}tv{\"o}s Lor\'and Research Network, Budapest, 1097, Hungary.}
\affil[4]{Corvinus University of Budapest, Budapest, 1093, Hungary.}
\affil[5]{Ume\aa\ University, Department of Economic history, SE-90187, Ume\aa\, Sweden.}
\affil[6]{Ume\aa\ University, Department of Geography, SE-90187, Ume\aa, Sweden.}
\affil[$\ast$]{To whom correspondence should be addressed: \href{email:r.di-clemente@exeter.ac.uk}{r.di-clemente@exeter.ac.uk}}
\affil[$\ddag$]{To whom correspondence should be addressed: \href{email:lengyel.balazs@krtk.hu}{lengyel.balazs@krtk.hu}}

\begin{abstract}
Core objectives of European common market integration are convergence and economic growth, but these are hampered by redundancy, and value chain asymmetries. The challenge is how to harmonize labor division to reach global competitiveness, meanwhile bridging productivity differences across the EU. We develop a bipartite network approach to trace pairwise co-specialization, by applying the Revealed Comparative Advantage method, within and between EU15 and Central and Eastern European (CEE). This approach assesses redundancies and division of labor in the EU at the level of industries and countries. We find significant co-specialization among CEE countries but a diverging specialization between EU15 and CEE. Productivity increases in those CEE industries that have co-specialized with other CEE countries after EU accession, while co-specialization across CEE and EU15 countries is less related to productivity growth. These results show that a division of sectoral specialization can lead to productivity convergence between EU15 and CEE countries.
\end{abstract}

\begin{document}

\flushbottom
\maketitle
\section*{Introduction}

The European integration has changed the production landscape of the continent and the development paths of its member states. The removal of European borders following the fall of the Iron Curtain and subsequent EU enlargement opened a free flow of goods and factor endowments between countries. In theory, such integration facilitates the convergence of countries that are at different development stages \cite{Solow1956}. In turn, others argue that the most and less developed countries and sectors might not equally tackle increasing competition within the common market \cite{Knox1960}. Some even claim that the emerging new division of labor creates asymmetric value-chain relations in the integrated macro-region, hampering economic and societal progress in less developed or accession countries \cite{rodriguez2018revenge,smith2002networks}.

A growing literature have therefore focused on the above duality of the EU integration process. Motivated by neoclassic growth theory, economists have investigated whether the unified market fosters economic growth of entrants and if so, whether convergence in income and productivity levels happen on the short or on the long run as a consequence of technological and knowledge spillovers towards less developed countries \cite{Palan2010,MIDELFART2003,Barro1991,Borsi2015,Kutan2007,Kutan2009,Petrakos2005,henrekson1997growth}. Recent studies focusing   on the Eastern enlargement find compelling support for the convergence hypothesis \cite{cabral2019europe, benczes2015european, nagy2022european}. Case studies on specific industries have looked into how re-location of production to accession countries and integration of their industries to European value chains influence development \cite{crestanello2011industrial,frigant2009modular}. Some have even emphasised convergence of economic structures as well by measuring structural similarity of trade and production between EU member countries \cite{Palan2010,Crespo2012,MIDELFART2003,Kallioras2010}.

However, labor division across nations in the global value chain is a natural process and a potential area of policy intervention to foster convergence in income by taking advantage of diverging specializations. Starting from the simple Ricardian model, comparative advantage has been considered a major source of national specialization into certain products or industries as international trade intensifies \cite{Balassa1965, leamer1984sources}. New economic geography (NEG) models on European integration suggest diverging structures of specialization as agglomeration forces of industry production grow stronger at lower barriers of trade \cite{krugman1996integration}. This process facilitates economic progress due to the accumulation of knowledge and skills through specialization in line with endogenous growth theory \cite{romer1987growth, marshall2009principles,archibugi1994aggregate}.  Dynamic models considering different stages of development suggest that countries tend to be more specialized the higher the levels of their income are \cite{imbs2003stages}, thereby questioning whether a deepened labor division actually can favor less developed countries when they integrate with more developed ones. 

More than a matter of specialization, also the type of specializations should make a difference for future development. Previous studies have identified that whether nations, for example, co-specialize into identical industries or, on the contrary, specialize into divergent structures is crucial for their development \cite{Hidalgo:2007aa, Hidalgo:2009aa}. This is because specialization into identical products can signal integration into value chains that can help income and productivity convergence \cite{frigant2009modular} but, at the same time, competition across countries as well \cite{fontagne2008specialization}. The latter would suggest that less developed countries can benefit less from co-specializing with more developed countries \cite{levchenko2012comparative,pop00008,saracco2015}. Yet, the strucutural properties of specialization has remained under-researched, especially in the context of convergence within the EU. 

In this paper, we offer a new  methodological framework that provides tools to understand both structure and dynamics, and also the consequences of co-specialization across countries. Our case is the European integration and we define a bipartite network, in which EU member states are linked to industries if they have revealed comparative advantage (RCA) in that particular industry \cite{Balassa1965} in terms of workforce. The unipartite representations of RCA networks have been proven useful to understand the development of countries \cite{Hidalgo:2007aa}. Here, instead of projecting the bipartite network into countries or sectors \cite{Zweig2011}, we keep the bipartite representation and investigate the abundance of co-specialization motifs \cite{Milo2002,Simmons2019}. 
This approach has already been used to understand systems in ecology \cite{Dormann2009}, trade networks \cite{Saracco2015a}, finance \cite{Gualdi2016} and scientific competition of countries \cite{Cimini2014} and  offers new tools to better understand the dynamics of co-specialization at multiple levels.

We define the co-specialization motif directly on the bipartite network as a subgraph that captures if two countries are specialized in the same industry. By measuring the statistical significance of this motif we can assess whether and where production structures are overlapping or diverging in the EU \cite{Saracco2016,Simmons2019,Clemente2020}. Here, we aim to disentangle co-specialization
within and across the groups of EU15 and CEE member states. We can also quantify these metrics on the industry-country level to create a metric that can predict their economic growth. Together these exercises allow us to discuss how labor division 
and co-specialization across countries influence convergence in productivity. 

Our measurement reveals no significant overlaps of comparative advantages across EU15 countries after 2000, which is probably due to their gradual integration. However, we find that CEE countries tend to be specialized in similar, if not identical, sectors. The number of co-specialization network motifs including both EU15 and CEE countries decrease after enlargement signalling a deeper division of production between EU15 and CEE. We do however find that productivity increases in those CEE industries that had no significant overlap in specialization before EU accession but experience increasing co-specialization with other CEE countries after entering the EU. In the meantime, co-specialization across EU15 and CEE countries contribute slightly positive, if at all significant, to productivity growth after accession. The findings refer to the role of labor division between EU15 and CEE that foster co-specialization of CEE industries and facilitate their integration and convergence in the common market.

\begin{figure*}[h!]
\centering\includegraphics[width=0.95\textwidth]{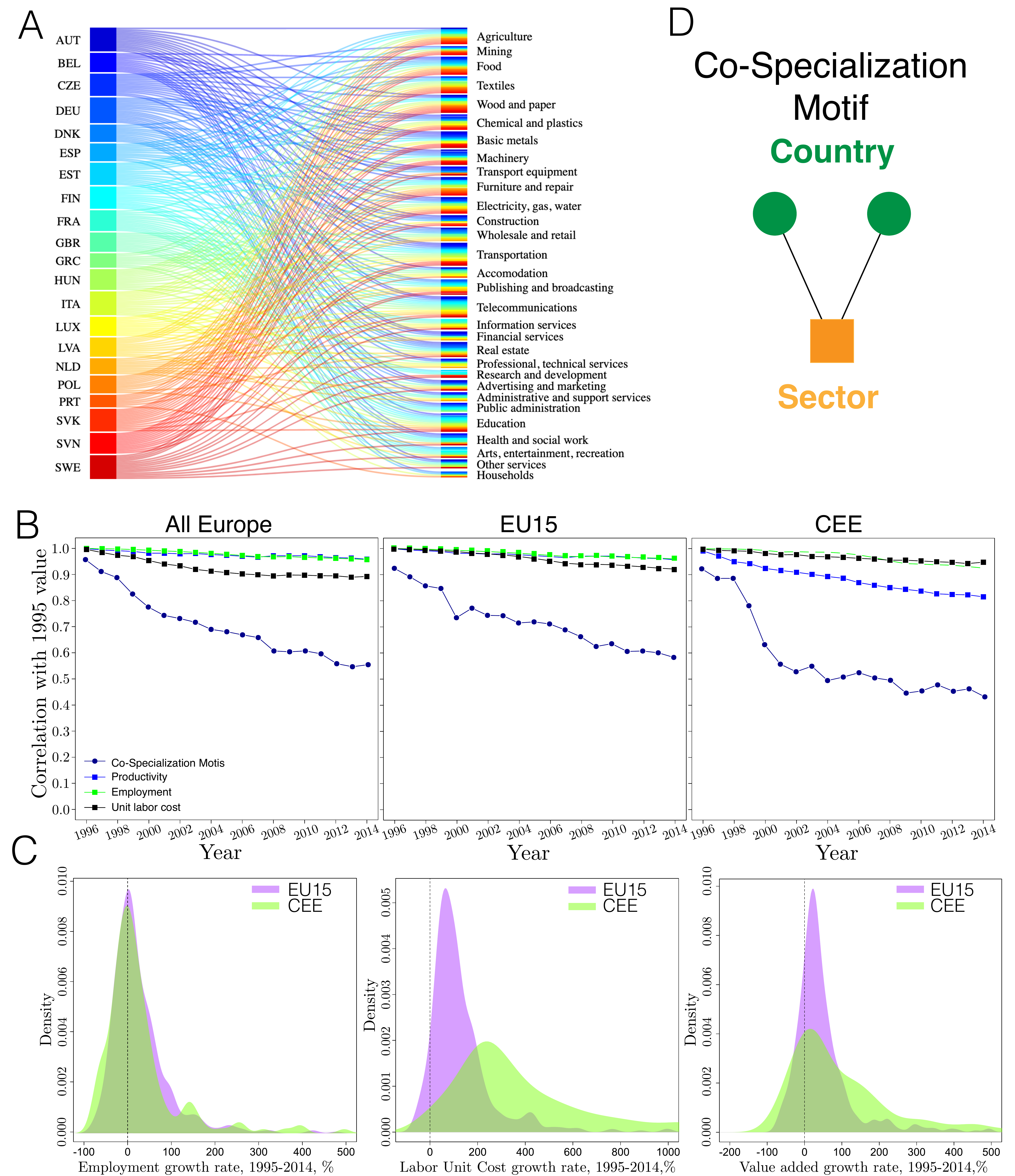}
\caption{
Variables and their dynamics. \textbf{(A)} Bipartite networks of European countries and industry sectors of 2005. \textbf{(B)} Correlation of country-industry variables with their 1995 value. The drop of coefficients of the co-specialization motif signal dynamic change whereas other measures suggest proportional growth across sectors. \textbf{(C)} Employment is stable in most EU15 and CEE industries and the number of workers grows extensively only in a limited number of sectors. Unit labor cost (average cost of employment per unit of output produced in million Euros) have grown more in CEE sectors than in EU15 sectors.  Value added per capita in million Euros has grown slowly in both country groups but the tail of the distribution is fatter in the CEE.  \textbf{(D)} Co-specialization motif across countries that have revealed comparative advantage in identical sectors.}
\label{fig_1}
\end{figure*}

\section*{Results}
\subsection*{Bipartite Network Approach to European Integration}

The core of our empirical approach is a bipartite, undirected and dynamic network between 21 European countries  $C=[1,\dots,c,\dots,21]$, $31$ industry sectors  $S=[1,\dots,s,\dots,31]$, and over $14$ years $t=[2000,\dots,t,\dots,2014]$ where the countries and the industries are the two types of nodes in each year (Figure \ref{fig_1}A).
The data for the analysis have been retrieved from the OECD webpage  and allow us to assess changes before the 2004 enlargement as well as evaluating the post-enlargement changes (see data description in Methods section).

We start from a rectangular matrix $\tilde{M}_t$ with dimensions $C$ and $S$ where each entry $\tilde{M}_{cs,t}$ is the number of employees of country $c$ in industry $s$ and year $t$.
The links between industries and countries are established by applying the Revealed Comparative Advantage (RCA) index as developed by Balassa \cite{Balassa1965} over the entries $\tilde{M}_{cs,t}$:

\begin{equation}
\label{eq_rca1}
	\mbox{RCA}_{cs,t}=\frac{\frac{\tilde{M}_{cs,t}}{\sum_{s'\in S}\tilde{M}_{cs',t}}}{\frac{\sum_{c'\in C}\tilde{M}_{c's,t}}{\sum_{c'\in C}\sum_{s'\in S}\tilde{M}_{c's',t}}}
\end{equation}

The entries of the adjacency matrix that represents the binary undirected bipartite industry-country network in year $t$ are defined by the rule:
\begin{equation}
\begin{cases}
\label{eq_rca2}
    M_{cs,t}=1 & \text{when } \mbox{RCA}_{cs,t}\geq 1\\
    M_{cs,t}=0 & \text{otherwise}
\end{cases}
\end{equation}

Eq. \ref{eq_rca2} indicates that the industry-country link is established if $\mbox{RCA}_{cs,t}$ is over the threshold $=1$.
In this case, country $c$ is considered to have comparative advantage in sector $s$ given the specialization in that sector in terms of employment\cite{frigant2009modular}. 
According to theory, such specialization leads to economic and technological progress \cite{Metcalfe2006} and thus international trade and economic development literatures have recently focused on the dynamics of RCA \cite{Hidalgo:2009aa,tacchella2012new}.

Our bipartite network approach allows us to further this research in two major methodological domains: 
\begin{itemize}
	\item dynamics of $\mbox{RCA}_{cs,t}$ in relation to sectors in other countries;
	\item the decomposition of $\mbox{RCA}_{cs,t}$ to groups of countries that are either of similar or of different stages of economic development
\end{itemize}

We propose that such methodological progress enables us to characterize the dynamics of co-specialization at multiple levels and thus infer the trends of labor division during European integration and also the role of different types of co-specialization in value added dynamics \cite{levchenko2012comparative}.

Our approach builds on network motifs that are patterns that typically consist few nodes of the network and the links between them \cite{Saracco2016,Milo2002}. In bipartite networks, motifs carry valuable information that is otherwise hidden in classic network analysis of unipartite networks projected from bipartite networks \cite{Saracco2017,Gualdi2016,VanLidthdeJeude2019}. Our focus is on the most simple co-specialization motif in this bipartite country-industry network defined by Eq. \ref{eq_co-spec} and illustrated in Figure \ref{fig_1}D.

The 'Co-Specialization Motif' $\mu^{co-spec}$ defines the pairs of countries in a given year that are both specialized in a given industry as $M_{cs,t}=M_{c's,t}=1$. $\mu^{co-spec}$ captures similarity between countries in having comparative advantage in certain industries \cite{Simmons2019}. To quantify the co-specialization in one or two sectors in a country with the same sectors in other countries, we count the number of $\mu_t^{co-spec}$ for every country-industry observation in each year:

\begin{equation}
	\mu_t^{co-spec}=\frac{1}{2}\sum_{s=1}^Su_{s,t}(u_{s,t}-1)
	\label{eq_co-spec}
\end{equation}

where, $u_{s,t}=\sum_{c\in C} M_{cs,t}$ is a vector of length $C$ , giving the degree of each columns, it is representing how an industry sector is ubiquitous.
Certainly, more complex motifs\cite{Simmons2019, Clemente2020} can be quantified to unfold various types of RCA relations across countries and industries offering tools to analyse the intensity of inter-relatedness of specialization.

The indicator introduced in Equation \ref{eq_co-spec} enables us to capture the dynamics of RCA relations between sectors and across countries and provide new information on economic restructuring during European integration. To illustrate this in Figure \ref{fig_1}B, we correlate $\mu_t^{co-spec}$ with its 1995 values and do the same exercise of employment, unit labor costs, and value added in sectors for the entire EU and decomposed to EU15 and CEE countries. Stable correlation coefficients of employment and value added in EU15 industries denote a rather stable structure of proportional growth. Together with the observation in Figure \ref{fig_1}C we find that most EU15 industries have not or have only slightly grown over 1995-2014. There is however a small deviation in terms of unit labor costs starting from 2004, the entry year of the accession countries denoting a restructuring shock of accession to EU15 wages. In the CEE countries, the growth of employment and unit labor costs are proportional despite the immense growth rate of the latter. The correlation coefficient of value added in CEE decreases monotonically but remains high as well denoting that ranks of CEE sectors in terms of productivity have slightly changed over the period.

In contrast, the correlation coefficient of $\mu_t^{co-spec}$ is decreasing monotonically and sharply suggesting that RCA relations are changing much more over time than other indicators frequently used to understand EU integration (see Figure \ref{fig_1}C). We find that co-specialization of the CEE countries were experiencing great internal turbulences before the entry, but less so after the entry in 2004. See Materials and Methods for further explanation of our bipartite network approach.

\subsection*{Co-specialization and Labor Division Revealed by Motif Significance}

To assess whether network motifs carry significant information about co-specialization of countries into industries, we compare the occurrence of observed motifs to a null model for each year using the Bipartite Configuration Model \cite{Saracco2016}. This is done by calculating the z-score of $\mu_{cs,t}^{co-spec}$ by:

\begin{equation}
	z_{cs,t}^\mu=\frac{\mu_{cs,t}^{co-spec}-{\langle \mu\rangle}}{\sigma_{\mu}}
\label{eq_zscore}
\end{equation}

where $\langle \mu\rangle$ is the average of $\mu_{cs,t}^{co-spec}$ in a set of 10,000 random matrices that persevere the average degree sequence of nodes in $M_{cs,t}$ and  $\sigma_{\mu}$ is the standard deviation of these randomized distributions. Values of $z_{cs,t}^\mu$ around zero mean that the observed motif distribution does not differ significantly from the null model. Positive high values capture significant co-specialization; while negative low values denote dissimilar RCA structures across countries. For a detailed description of the Bipartite Configuration Model, see Materials and Methods.

\begin{figure*}[h!]
\centering\includegraphics[width=0.80\textwidth]{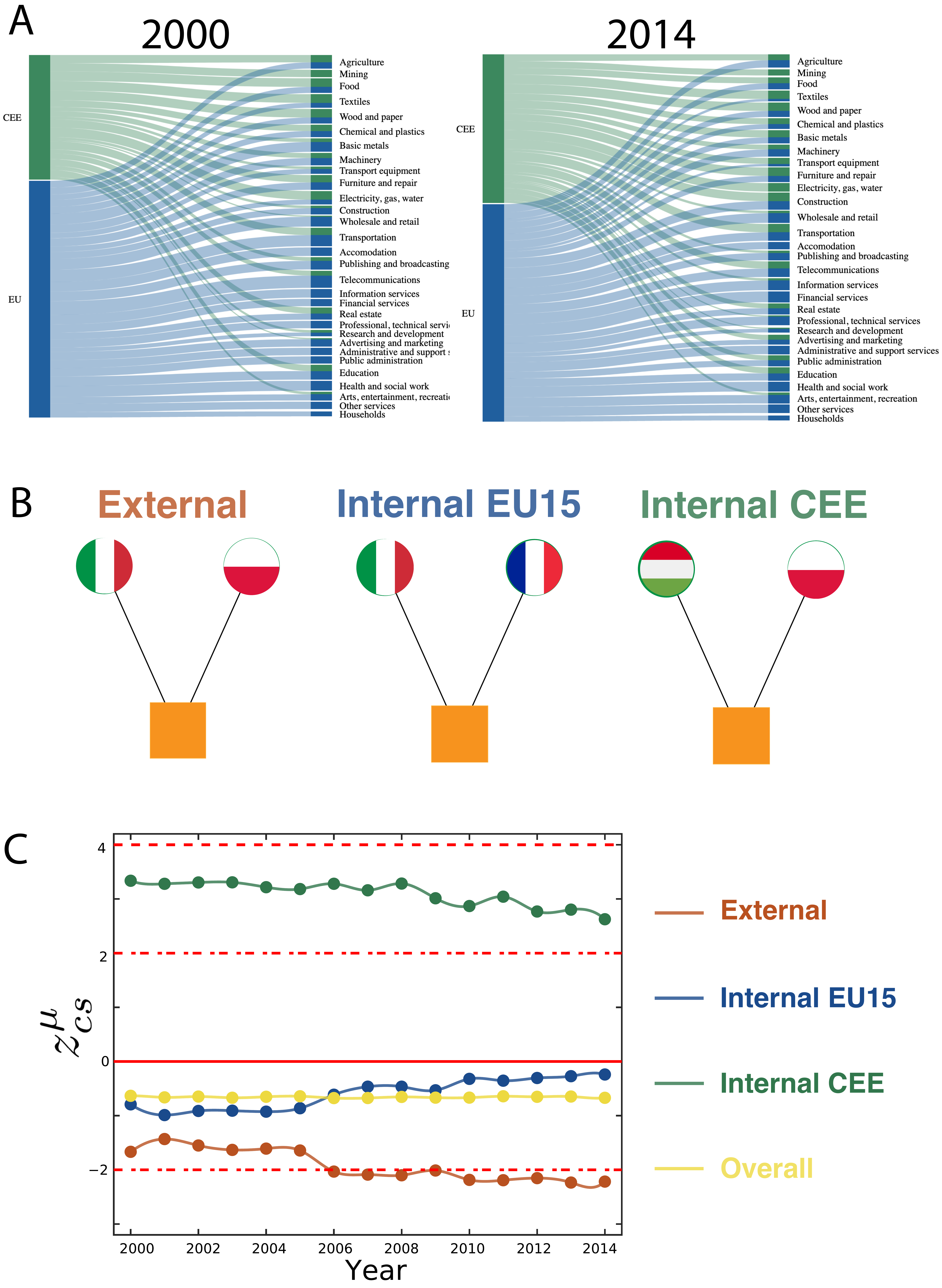}

\caption{
Co-specialization within and between country groups. \textbf{(A)} The share of CEE countries in the bipartite RCA network has grown from 2000 to 2014. \textbf{(B)} Illustration of Internal and External co-specialization motifs. The exemplar External Motif is illustrated with a link between Italy (EU15) and Poland (CEE); the Internal EU15 is represented by a link between Italy and France and the Internal CEE with a link between Hungary and Poland.  \textbf{(C)} Trends of co-specialization motifs compared to a null model of co-specialization, quantified by the z-score, illustrate that CEE countries co-specialize in identical industries but EU15 countries do not. The negative z-score of External co-specialization motifs denote an intensifying labor-division between EU15 and CEE that becomes significantly different from the null model after 2006.}
\label{fig_2}
\end{figure*}

Tracing $z_{cs,t}^\mu$ over time allows us to infer on co-specialization trends during the EU enlargement process. We label countries in the bipartite network CEE (accession countries) and EU15. Figure \ref{fig_2}A demonstrates considerable dynamics between 2000 and 2014, in which CEE industries have a growing share of RCA links (for example in Basic Metals, Transport Equipment or Telecommunications). To quantify co-specialization dynamics within and across these country groups, we distinguish three cases of co-specialization motif that are illustrated in Figure 2B. Internal EU15 $\mu_{c\in \mathrm{EU15},t}$ denotes motifs with two EU15 countries specialized in the same sector while Internal CEE $\mu_{c\in \mathrm{CEE},t}$ captures specialization of two CEE countries in the same sector. External $\mu_{\mathrm{EXT},t}$ stands for motifs that capture specialization of one EU15 country and one CEE country in the same industry. We can decompose the abundance of the co-specialization motif in:

\begin{equation}
\begin{split}
	\label{eq_ext}
	\mu_{\mathrm{EXT},t}=\mu_{t}-(\mu_{c \in  \mathrm{EU15},t}+\mu_{c \in  \mathrm{CEE},t})=\\
	=\sum_{s}\sum_{c'\in \mathrm{EU15}}\sum_{c''\in \mathrm{CEE}} M_{c's,t} M_{c''s,t}.
\end{split}
\end{equation}

Figure \ref{fig_2}C reports on $z_{cs,t}^\mu$ trends over EU enlargement. The z-score measured from the full matrix containing all countries is stable around $-0.5$. Restricting the countries to EU15 produces similar z-scores that even converge to zero. These findings mean that co-specialization across all EU countries and across EU15 countries is not significantly different from the null model.

However, both Internal CEE co-specialization and External co-specialization are significantly different from the null model and reveal interesting trends. Internal CEE has high positive values that decrease over time, especially after 2008. 
There are two potential  reasons for the drop of co-specialization after the financial crisis. Firstly, the unequal impact of the crisis across member states, and secondly, the different paths of recovery during its aftermath \cite{kolev2012impact, HRISTOV2012569, szekely2009economic}. External co-specialization z-scores have low negative values meaning that EU15 and CEE countries tend to have comparative advantages in distinct industries. The z-score becomes significant in 2006 signalling that EU accession has been followed by a stronger division of production across old and new member states. Supplementary Material 1 contains the z-score trends decomposed to sectors, which reveal different dynamics. For example, there is a significant co-specialization in Primary production of the CEE countries but this sector has a non-overlapping RCA structure in EU15 countries. We find that EU15 and CEE countries tend to not specialize in the same service sectors.

\subsection{Impact on productivity}

In this section, we give an example of the industry-country level application of our framework. By rewriting Equation \ref{eq_co-spec}, it is possible to study the co-specialization of a sector in a country. The significance level of the co-specialization motif of industry $s$ in country $c$ changes in case the of number of industries in other countries that the given sector in a given country is co-specialized with increases or decreases. This enables us to analyse the relation between the dynamics of co-specialization motifs and productivity of industry-country pairs. 

The industry-country level co-specialization is defined by:

\begin{equation}
	\mu_{cs,t}^{co-spec}=(u_{s,t}-1)M_{cs,t}.
\end{equation}

Following the logic of Equation \ref{eq_ext}, we decompose $\mu_{cs,t}^{co-spec}$ to $\mu_{cs,t}^{\mathrm{EXT}}$, $\mu_{cs,t}^{\mathrm{CEE}}$, $\mu_{cs,t}^{\mathrm{EU15}}$. For each country groups CEE and EU15, we define the INTERNAL indicator that includes co-specialization within CEE and EU15 groups of countries ($\mu_{cs,t}^{\mathrm{CEE}}$,$\mu_{cs,t}^{\mathrm{EU15}}$). The EXTERNAL indicator includes co-specialization across CEE and EU15 groups ($\mu_{cs,t}^{\mathrm{EXT}}$). From now on, we refer to $\mu_{cs,t}^{co-spec}$ as the OVERALL Motif of co-specialization.

\begin{figure*}[h!]
\centering\includegraphics[width=0.99\textwidth]{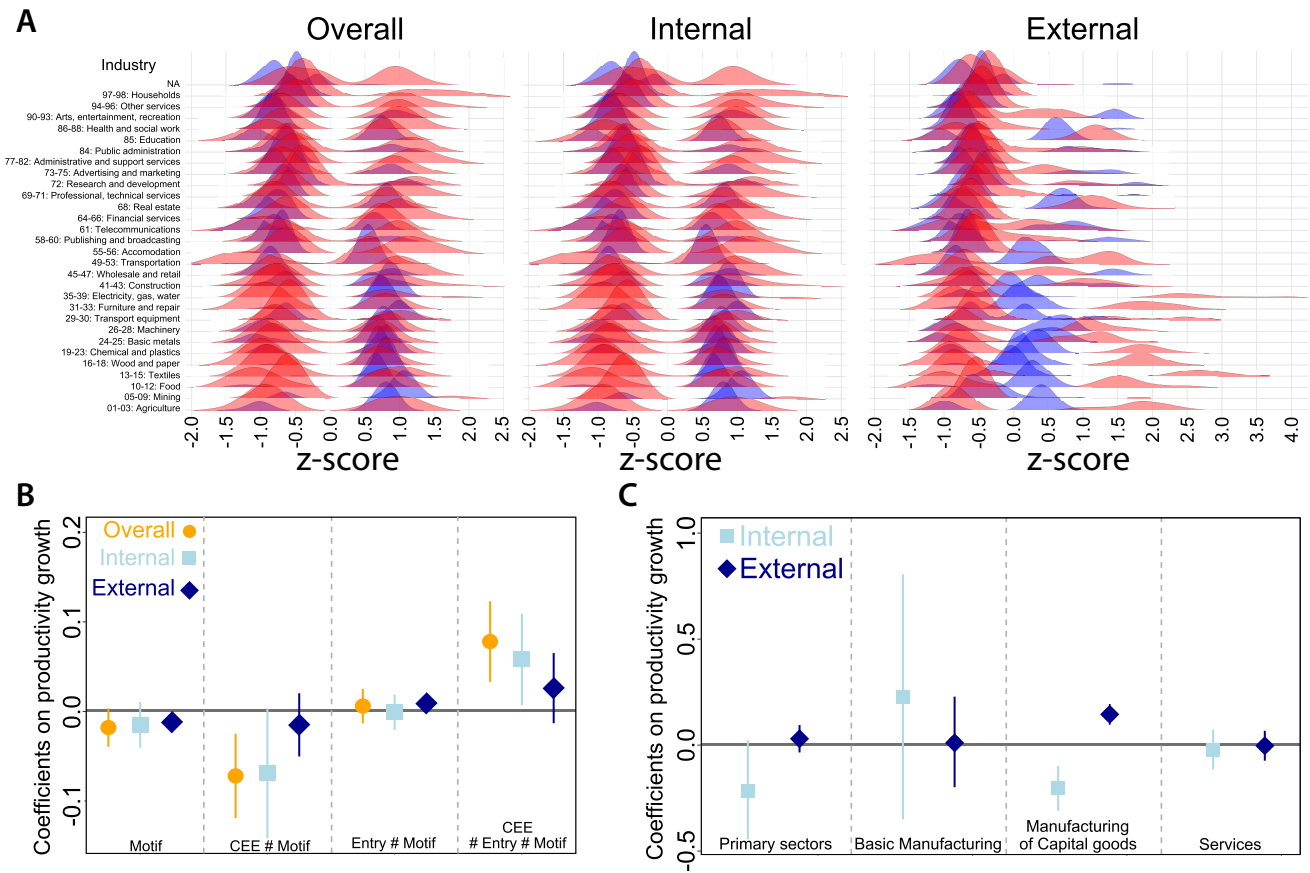}
\caption{
Co-specialization motifs and value added growth.  \textbf{(A)}. z-scores of co-specialization are plotted over the 2000-2014 period. Blue depicts z-scores in CEE industries and red is z-scores in EU15 industries. \textbf{(B)}. The role of co-specialization motifs and their effect in CEE countries on value added growth after entering the EU. \textbf{(C)}. The role of External and Internal motifs for value added growth by industries during EU integration.}
\label{fig_3}
\end{figure*}

Figure \ref{fig_3}A illustrates the distributions of OVERALL, INTERNAL and EXTERNAL Motifs by sectors.
The figure shows that CEE countries (in blue) tend to co-specialize in manufacturing sectors while EU15 countries (in red) tend to co-specialize in service sectors. Decomposing OVERALL similarity to EXTERNAL and INTERNAL, we find that CEE countries co-specialize with other CEE countries in manufacturing sectors to a high degree and EU15 countries co-specialize with other EU15 countries in services but in a slightly lower degree. Co-specialization across EU15 and CEE countries mostly appears for CEE industries and mostly in manufacturing industries. However, in few cases this is highly significant for EU15 industries as well.

To estimate the impact of the above described changes in motifs on the performance (value added per full-time equivalents) of each industry in every country, we have employed a difference-in-differences \cite{pavlinek2002transformation} approach in a fixed effect panel model for the period 2000-2014 that is specified by:

\begin{equation}
	\label{eq_reg}
	y_{i,t}=\beta'X_{i,t-1}+\gamma'Z_{i,t-1}+\lambda_i+\epsilon_{i,t},
\end{equation}

where $y$ denotes value added per capita, $t$ denotes one-year intervals from 2000 to 2014, $i$ denotes the pair sector-country $i=(c,s)$ , $X$ represents a vector of co-specialization variables, $\lambda_i$ denotes the industry-country fixed effects, $Z$ stands for a set of control variables and $\epsilon$ is the case- and time-specific error term. The definition, descriptive statistics and correlation values of the indicators can be found in Supplementary Material 2 while Supplementary Material 3 explains the motivation for this specification in detail.

This regression technique allows us to assess how changes in motifs influence value added in relation to (i) the entry of CEE in the EU, and (ii) whether such changes occur in CEE or in EU15. This is important to distinguish because some capital adjustment is likely to occur as new sites of production and new markets become available within the common market. Hence, a dummy variable (Entry) which equals one for the period 2004 and onwards were created to capture the role of the enlargement as such. Moreover, another dummy variable (CEE) that equals 1 if a given part of the CEE countries joining the EU in 2004 (i.e., Czech Republic, Estonia, Hungary, Latvia, Poland, Slovakia or Slovenia). Finally, two controllers capturing the size of the industry (EMP) and gross capital formation (GFC) are added since value added might be driven by labor saving technologies. All continuous variables are standardized to ease interpretation.  

The results from the regressions on the full models that include interaction-terms are displayed in Figure \ref{fig_3}B. The first results on the OVERALL Motif indicate that the main effect of co-specialization does not significantly influence value added growth. In fact, there is a negative overall influence of increasing EU-wide co-specialization in the accession countries before the entry, but that turns positive after the entry. In other words, increasing post-entry co-specialization seems to primarily benefit the CEE countries. This would indicate that, if anything, an increasing convergence in overall EU co-specialization is beneficial for growth. 

However, when decomposing this co-specialization into the INTERNAL and EXTERNAL terms (CEE and EU15 respectively), our results indicate a somewhat more complex pattern. First, while the main effect of increasing internal co-specialization is insignificant, the corresponding estimate on increasing external co-specialization is significantly negative. Due to the corresponding interaction effects this implies that this negative association mainly is identified in the EU15 before the entry and then slightly moderated after the entry. Concerning the CEE countries, our findings on internal co-specialization suggest that before the entry increasing co-specialization within CEE hampers growth, while after the entry increasing internal co-specialization in the CEE explain the overall positive association with growth in the CEE. Hence, it is not convergence within the EU that contribute to growth, but rather convergence within the accession countries after the entry. As is displayed in Figure \ref{fig_3}C where we decompose INTERNAL and EXTERNAL terms in different groups of sectors, the impact on productivity growth is positively associated with internal co-specialization in basic manufacturing and external co-specialization in manufacturing in capital goods. While division of labour between CEE and EU15 in basic manufacturing largely impact on productivity growth, capital goods is not affected by the entry as such. Regression tables and robustness checks are presented in Supplementary Material 4.

\section*{Discussion}
Altogether, our findings imply a deeper division of labor across diversified EU15 and specialized CEE countries. This division follows from the principle of comparative advantages, where EU15 and CEE countries specialize in producing goods and services at a lower opportunity cost to trade. While NEG models assume that as the EU become more integrated, economic activity become more regionally concentrated and specialized \cite{krugman1996integration}, recent empirical studies fail to find consistent support for this argument \cite{cutrini2022eu}. Our findings on the structural changes of specialization indicate that the convergence effects of this evolution depends on the pattern of specialization. Thus, with whom´s structure the accession countries converge towards. The regression results confirm that productivity increases in those CEE industries that converge in specialization with other CEE countries, not with other EU members, while co-specialization across CEE and EU15 countries is less related to productivity. If anything, co-specialization across EU-15 and CEE is more beneficial for the EU15 which suggest difficulties for CEE countries to specialize in similar activities as EU15 countries after the accession. 

Since these two country groups are known to be at different stages of development \cite{du2010inter} this indicates two different processes of EU integration. On the one hand, integration has made offshoring of certain stages of the value chain possible which has increased internal CEE co-specialization and growth. This may also reflect the fact that countries that have similar production endowments tend to trade with similar but related products to utilize returns to scale by specializing in different tasks \cite{grossman2012task}. On the other hand, in contrast to the inter-sectoral convergence previously identified for Western European countries as lagging-behind countries has shifted from industrialized to service economies \cite{Palan2010}, our findings suggest that the differences between EU15 and the CEE countries instead has increased over time. This, in turn, suggests greater differences of sector specializations and thus a persistent and even intensifying labor division between the EU15 and the CEE countries. Based on the decomposed estimates this could be explained by the fact that basic manufacturing is expanding across all CEE countries after the accession which increases both co-specialization within CEE and productivity. Given the negative estimate for external co-specialization this pattern is partly driven by relocation of such activities from the EU15. Finally, our findings suggest that the remaining co-specialization mainly benefit the EU15 that already have a productive edge as compared to the accession states. 

\section*{Conclusions}

While great emphasis has been put on tackling territorial disparities within the EU over time, persistent differences in employment and output still remains. Since the bulk of present Cohesion Policy funding is concentrated on less developed European countries and regions to help them catching-up, the main objective of this paper was to assess how structures of employment has changed in the CEE countries when entering the EU. While this is far from a new empirical assessment, the contribution of this paper was the use of a novel bipartite network approach to trace the significance of co-specialization in employment over the European integration 2000-2014. We present a new methodology to revealed comparative advantage interactions using the bipartite network approach that enables a more direct understanding of the countries' interaction.  Tehereby overcoming the limitations of monopartite approaches.

The methodological contributions offered in this paper regard the decomposition of significant network motifs. This, in turn, enables a quantification of the dynamics of co-specialization at the level of the whole EU, groups of countries, and industry-country pairs compared to more general specialization indices typically applied in the empirical literature. Since this method reveals distinct co-specialization trends across EU member countries, more detailed patterns of convergence and divergence within the EU can be revealed. For example, while co-specialization among EU15 member states is not significant, the pairwise specialization of CEE countries is significantly higher and the co-specialization across EU15 and CEE countries is significantly lower than random distributions would suggest.

The bulk of previous studies on convergence either advocates neoclassical growth theory or more endogenous mechanisms. The division of labour at different levels of development, and its relation to productivity growth identified in this paper, however underlines the interdependence between growth and trade suggested in the NEG models on "new trade theory" \cite{neary2009putting}. The increasing returns to scale associated with co-specialization furthermore underlines the ability to exploit economies of scale and market size suggested by endogenous growth theory \cite{helpman1987market}.

The division of labor between EU15 and CEE is persistent and is seemingly growing and therefore requires special policy focus.  Our findings highlight that the current EU policy on Smart Specialization \cite{McCann2015,Thissen2013,Capello2016} need to be applied  in new member states differently from old member states to reflect labor division that can support convergence in productivity. One of the risk factors this paper touches upon is the rapidly increasing unit labor costs in manufacturing over the past two decades in CEE economies. Although such a development entails social benefits in terms of higher incomes and potential for tax revenues, such a development may in the longer-run compromise competitiveness and reduce market share for manufacturing output.

While the advantage of the proposed method is its' ability to detect signals of co-specialization dynamics from simple data sources, the empirical approach provided here is not without limitations. We focused only on employment structures and its association with productivity, while the exact mechanisms of economic integration can be better captured with more detailed data. Future studies should therefore complement this approach with trade data to assert the potential evolving interdependence between specialization and trade.   

\section*{Methods}

\subsection*{Data}
Data has been retrieved from the OECD webpage\footnote{ \href{https://stats.oecd.org/Index.aspx?DataSetCode=STANI4_2016}{https://stats.oecd.org/Index.aspx?DataSetCode=STANI4\_2016} (last checked 09/2021)}. Country grouping is presented in Table \ref{tab_1} and sectors classes are presented in Table \ref{tab_2}.

\begin{table*}[ht]
\caption{\label{tab_1}Country Groups.}
\tabcolsep=0pt%%
\begin{tabular*}{\textwidth}{@{\extracolsep{\fill}}lll@{\extracolsep{\fill}}}
\toprule%
Group Name & Countries ISO CODE \\
\hline
CEE & CZE, EST, HUN, LVA, POL, SVK, SVN \\
EU15 & AUT, BEL, DEU, DNK, ESP, FIN, FRA, GBR, GRC, ITA, LUX, NDL, PRT, SWE\\
\hline
\end{tabular*}
\end{table*}

\begin{table*}[ht]
\caption{\label{tab_2}Industries Sectors.}
\tabcolsep=0pt%%
\begin{tabular*}{\textwidth}{@{\extracolsep{\fill}}lll@{\extracolsep{\fill}}}
\toprule%
Sectors Name & Codes STANI4 2016 \\
\hline
Primary production & D01T03, D05T09, D10T12, D13T15, D16T18 \\
Basic manufacturing & D19T23, D24T25 \\
Manufacturing of capital goods &D26T28, D29T30, D31T33 \\
Infrastructure & D35T39, D41T43 \\
Retail & D45T47, D49T53 \\
Services & DD55T56, D58, T60, D61, D62T63, D64T66, D68, D69T71, D72, D73T75, D77T82 \\
Personal services & D84, D85, D86T88, D90T93, D94T96, D97T98, D99 \\
\hline
\end{tabular*}
\end{table*}

\subsection*{Mathematical Appendix}

\subsubsection*{Notation \& Bipartite Configuration Model}
In this section we will follow the definitions provided by \cite{Saracco2015a,Simmons2019,Clemente2020}. All the formulas are defined for the analysis of given year $t$. 
The rows of $M_{cs,t}$ correspond to Countries $C=[1,\dots,c,\dots,21]$, the columns correspond to Industry Sectors $S=[1,\dots,s,\dots,31]$. $u_{s,t}=\sum_c^CM_{cs,t}$ is a vector of length $C$, giving the degree of each columns, it represents the ubiquity of an industry sector, whereas $d_{c,t}=\sum_s^CM_{cs,t}$ is a vector of length $S$, giving the degree of each row and quantifies how diversified a country is.

The Bipartite Configuration Model (BiCM) is a null model for bipartite, undirected, binary networks,  that is able to generate a grandcanonical ensemble of networks as defined by \cite{Saracco2015a,Saracco2016}.
BiCM generates a network ensemble where each matrices has constrained the number of links for each node, on both layers (i.e. $d_{c,t}$ and $u_{s,t}$) to match, on average, the observed one.
Each network $\mathbf{M}$ in such ensemble is assigned a probability coefficient:
 
\begin{equation}
P(\mathbf{M}|\vec{x}, \vec{y})=\prod_cx_c^{d_c(\mathbf{M})}\prod_sy_s^{u_s(\mathbf{M})}\prod_{c, s}(1+x_cy_s)^{-1},
\end{equation}

with $x_c$ and $y_s$ the Lagrange multipliers associated to the constrained degrees, of countries and sectors respectively.

The constrains of the ensemble average values of countries' and sectors' degree allow us to calculate the probability that a link exists between country $c$ and industry sector $s$ independently of the other links:

\begin{equation}
p_{cs}=\frac{x_cy_s}{1+x_cy_s}.
\label{prob}
\end{equation}

By solving the system of $C+S$ equations it is possible to determine the numerical values of the unknown parameters $\vec{x}$ and $\vec{y}$, which constrains the ensemble average values of countries' diversification and sectors' ubiquity to match the real values, $\langle d_c\rangle=d_c^*,\:c=1\dots C$ and $\langle u_s\rangle=u_s^*,\:s=1\dots s$.

 Where $\{d_c^*\}_{c=1}^C$ and $\{u_s^*\}_{s=1}^S$ are the real degree sequence of countries, and industry sectors respectively, and $\langle \cdot\rangle$ represents the ensemble average of a given quantity, over the ensemble measure defined by Eq.\ref{prob} - as $\langle d_c\rangle=\sum_sp_{cs}$ and $\langle u_s\rangle=\sum_cp_{cs}$. Indicated with an asterisk, ``$\ast$" are the parameters that satisfy the systems. 

$\mathcal{Z}_t$ is the matrix of dimension $(CxC)$, that represents the projection of $M_{cs,t}$. Each entry $\mathcal{Z}_{cc',t}$ counts the number of industry sectors in common between the countries $c$ and $c'$. It is defined as:
\begin{equation}
	\mathcal{Z}_{cc',t}=\sum_{s=1}^SM_{cs,t}M_{c's,t}.
\end{equation}

\subsubsection*{Co-specialization Motif}

The degree of an industry sector in the biadjacency matrix represents the number of countries that the industry in the focal country co-specialize with. To count the abundance of the co-specialization motif we need to count all the possible couple that can be generated within a set of $u_s$ countries. Hence, we have ${u_s \choose 2}=\frac{1}{2}u_s(u_s-1)$ possibilities to pick two countries that together with $s$ will form the motif. Following Eq. \ref{eq_co-spec} we can obtain the overall abundance of the co-specialization motif for the entire matrix $M$.

\subsubsection*{Research context of BiCM}
The BiCM it is an extension of the Binary Configuration Model to Bipartite networks \cite{Cimini2019}, which both are null model that uses the framework of Exponential Random Graphs (ERG) \cite{Holland1981}.
The BiCM is of general applicability, and it has been proven valuable in several multidisciplinary studies to quantify the structural modification on the bipartite network over time.
For example in the bipartite network representation of the World Trade Web (i.e. network layers Country / Exported Products) \cite{Saracco2015a,Saracco2016}, BiCM was able to define benchmark to highlight meaningful correlations between countries and products. The method detected early-warning signals of the changing WTW topology by analyzing the motifs of products' similarities in emerging economies. BiCM makes the validation of the one-mode projection of the bipartite WTW possible. This permits to recognize country pair-wise similarities in industrial systems and to detect statistically significant signals of export specialization dynamics developing from basic to more sophisticated products \cite{straka2017,Saracco2017}.
In ecology, BiCM was used to assess the interaction between the species in mutual networks and to show the nested shape of this interaction \cite{Payrato-Borras2019}.
Finally, BiCM applied on the bipartite representation of stocks ownership by financial institutions showed that increasing portfolio similarities between financial institutions before the 2008 financial crisis enhanced the systemic risk from fire sales liquidation and could be used to forecast market crashes and bubbles \cite{Gualdi2016}.

\section*{Funding}
B.L. acknowledges financial support from the Hungarian Scientific Research Fund (OTKA K-129207). R.E. acknowledge the financial support from the Marianne and Marcus Wallenberg foundation (2017.0042). All authors benefited from the support of the research program 'Mobility, Transformation and Regional Growth' at Ume\aa\ University. 

\section*{Author contributions statement}
R.D.C. developed and performed the mathematical framework of the Bipartite network analysis. B.L. analyzed the Industry dynamics in EU. L.-F.A. and R.E. defined and performed the econometric model. R.D.C., B.L., L.-F.A. and R.E. designed the study and performed the research. All the figures, plots and visualization have been created by the authors. All the authors wrote the paper and gave the final approval to the publication.

\section*{Data availability}
Data has been retrieved from the OECD webpage\footnote{ \href{https://stats.oecd.org/Index.aspx?DataSetCode=STANI4_2016}{https://stats.oecd.org/Index.aspx?DataSetCode=STANI4\_2016} (last checked Oct/2022)}.

\section*{Code availability}

Bipartite Configuration Model code can be found here: \href{https://bicm.readthedocs.io/en/latest/}{https://bicm.readthedocs.io/en/latest/} (last checked Oct./2022) 

Motifs and position motifs code can be found here:
\href{https://github.com/SimmonsBI/bmotif}{https://github.com/SimmonsBI/bmotif} (last checked Oct./2022).

\newpage
\section*{Supplementary Material}
\renewcommand{\figurename}{Figure}
\renewcommand{\thefigure}{S\arabic{figure}}
\setcounter{figure}{0}
\renewcommand{\tablename}{Table}
\renewcommand{\thetable}{S\arabic{table}}
\setcounter{table}{0}

\section{Supplementary Material 1: Z-score trends of sectors}
\begin{figure}[!htb]
\centering\includegraphics[width=0.99\textwidth]{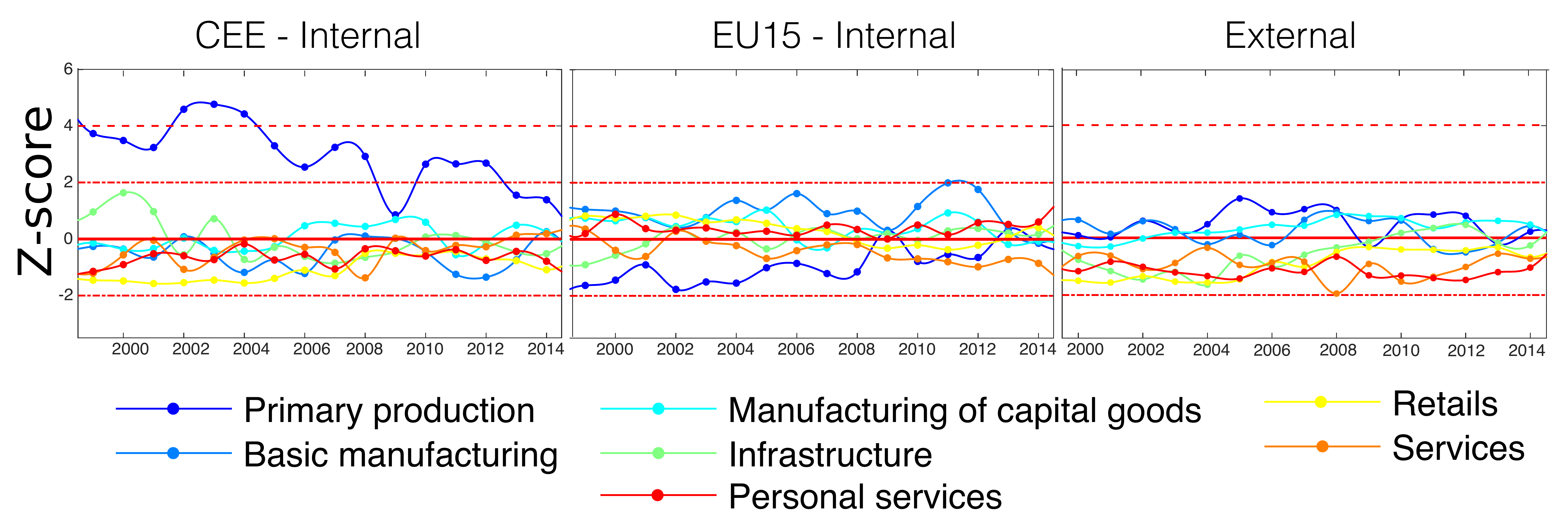}
\caption{Z-scores of co-specialization motifs decomposed into main sectors.}
\label{fig_S1}
\end{figure}

\newpage
\section{Supplementary Material 2: Definitions, descriptive and correlations of country-industry variables}

\begin{table}[ht]
\centering
\caption{Definition of regression variables}
\resizebox{0.90\textwidth}{!}{
\begin{tabular}{ |c|c| } 
 \hline
Variable & Definition \\ 
\hline
Overall & z-score of co-specialization motifs without country limitations (standardized)\\
Internal & z-score of co-specialization motifs within groups of EU15 or CEE countries (standardized)\\ 
External & z-score of co-specialization motifs across groups of EU15 and CEE countries (standardized)\\ 
Entry & Dummy=1 if year >= 2004  \\ 
CEE & Central and Eastern European Countries \\ 
Recession & Dummy=1 if year = 2008 or 2009 \\ 
GFC & Gross capital formation (standardized) \\ 
EMP & Employment (standardized) \\ 
 \hline
\end{tabular}}
\end{table}

\begin{table}[ht]
\centering
\caption{Summary statistics of regression variables}
\resizebox{0.50\textwidth}{!}{
\begin{tabular}{ |c|c|c|c|c|c| } 
 \hline
Variable & Obs & Mean & St.Dev. & Min & Max \\ 
\hline
Overall & 9,555 & 0.001 & 0.999 & -2.156 & 2.852 \\ 
Internal & 9,555 & -0.001 & 0.993 & -1.691 & 3.912 \\ 
External & 9,555 & 0.002 & 0.991 & -1.863 & 6.016 \\ 
Entry & 11,160 & 0.733 & 0.442 & 0 & 1 \\ 
CEE & 11,160 & 0.292 & 0.454 & 0 & 1 \\ 
Recession & 11,160 & 0.067 & 0.249 & 0 & 1 \\ 
GFC & 8,572 & 0.261 & 0.772 & -8.309 & 6.112 \\ 
EMP & 10,977 & 0.004 & 1.015 & -0.496 & 9.935 \\ 
 \hline
\end{tabular}}
\end{table}

\begin{table}[ht]
\centering
\caption{Pair-wise Pearson correlation of regression variables}
\resizebox{0.70\textwidth}{!}{
\begin{tabular}{ c|c|ccccccc } 
 \hline
\# & Variable & 1 & 2 & 3 & 4 & 5 & 6 & 7\\ 
\hline
1 & Overall &  & & & & & & \\ 
2 & Internal &  0.888 & & & & & & \\ 
3 & External &  0.709 & 0.345 & & & & & \\ 
4 & Entry &  0.002 & 0.011 & -0.009 & & & & \\ 
5 & CEE &  0.002 & 0.102 & -0.023 & -0.000 & & & \\ 
6 & Recession &  0.001 & 0.003 & -0.003 & 0.161 & -0.000 & &\\ 
7 & GFC &  0.023 & 0.031 & -0.014 & 0.440 & -0.057 & 0.078 &\\ 
8 & EMP &  0.147 & 0.151 & 0.011 & 0.009 & -0.174 & 0.004 & 0.033\\ 
 \hline
\end{tabular}}
\end{table}

\newpage
\section{Supplementary Material 3: Regression motivation}

Fixed effect (FE) panel regression models are applied to analyze the statistical relation between the motifs and performance in industry-countries. 

The rationale for using this type of model is that it allows us to explicitly control for unobserved (time-invariant) heterogeneity across industry-countries (such as norms and relative location within the EU or other aspects not captured by the controllers or by the definition of industries), which in itself may help reduce the impact of endogeneity. This is highly relevant in the European context due to the great variety of countries in terms of size, population, and economic structure which may influence the impact of new members differently. Moreover, owing to the within estimator that characterizes the fixed-effect model, it conditions how a change in motifs influences a change in value added over time. Compared to a pooled ordinary least squares (OLS) model where the between effect is emphasized (i.e., differences between cases), this approach emphasizes the dynamic relationship between our explanatory variables and the dependent variable over time. 

The fixed-effect approach thus permits us to model changes in value added in one industry-country in relation to changes in motifs over time in that particular unit. A Hausman test comparing a random-effect model with the fixed effect model also confirms that the fixed-effect model is more efficient. We reduce the risk of reversed causality influencing the results by having all explanatory variables measured the year before the dependent variable. Apart from the case-specific fixed-effects, all models include a dummy capturing the recession 2008-2009, and cluster-robust standard errors at the country-level to control for the fact that the dynamics of economic activities within a given country might be more similar than between countries.

Since the dummy CEE is a time-invariant variable, it cannot be estimated within a FE-setting. Therefore, each model has been estimated in a two-step procedure. First, the main effects of each motif, Entry and the control variables are estimated to assess the general role of motifs and enlargement. Second, we include interaction effects between the motifs, Entry and CEE to explicitly assess whether changes in motifs in the CEE countries vs EU15 in relation to the enlargement influence performance. 

Finally, all continuous variables are standardized (re-scaled to have a mean of zero and a standard deviation of one) to ease interpretation. 

\newpage
\section{Supplementary Material 4: Regression table}

\begin{table}[ht!]
\centering
\caption{\label{tab_s1}Fixed effect models on the impact of similarity (Sim) and overlap employment motifs and the entry of CEE countries in the EU on per capita value added per country-sector 2000-2014. Models 1-4 are on overall motifs and 5-8 separate between internal and external motifs. Cluster-robust SE:s at country level within brackets. Significant at $10\%$ (*), $5\%$ (**) and $1\%$ (***) confidence intervals.}
\resizebox{\textwidth}{!}{
\begin{tabular}{lcccc|cccc}
\hline\hline
&(1)&(2)&(3)&(4)&(5)&(6)&(7)&(8)\\
&All&All&All&All&Primary&Basic Manu&Capital Manu&Services\\
\hline\hline
\textit{Main}&&&&&&&&\\
Overall&-0.020**&-0.018&&&&&&\\
&(0.008)&(0.011)&&&&&&\\
Internal&&&-0.027**&-0.015&-0.211*&0.228&-0.204*&-0.021\\
&&&(0.012)&(0.013)&(0.119)&(0.295)&(0.054)&(0.048)\\
External&&&-0.007&-0.012*&0.030&0.015& 0.145*&-0.003\\
&&&(0.007)&(0.006)&(0.033)&(0.109)&(0.025)&(0.036)\\
\textit{Interactions}&&&&&&&&\\
CEE \# Overall&&-0.072**&&&&&&\\
&&(0.024)&&&&&&\\
Entry \# Overall&&0.006&&&&&&\\
&&(0.007)&&&&&&\\
CEE \# Entry \# Overall&&0.078***&&&&&&\\
&&(0.023)&&&&&&\\
CEE \# Internal&&&&-0.069*&0.137&-1.050*&0.281***&0.073\\
&&&&(0.037)&(0.136)&(0.561)&(0.086)&(0.081)\\
Entry \# Internal&&&&-0.001&0.080&-0.087&0.035&0.008\\
&&&&(0.010)&(0.173)&(0.280)&(0.099)&(0.026)\\
CEE \# Entry \# Internal&&&&0.058**&-0.009&0.864*&-0.115&-0.129**\\
&&&&(0.026)&(0.187)&(0.488)&(0.124)&(0.058)\\
CEE \# External&&&&-0.015&0.196&1.019&-0.278**&-0.115\\
&&&&(0.018)&(0.193)&(0.665)&(0.120)&(0.081)\\
Entry \# External&&&&0.009&0.-0.017&0.016&-0.041&0.081\\
&&&&(0.006)&(0.045)&(0.091)&(0.029)&(0.038)\\
CEE \# Entry \# External&&&&0.026&-0.377&-1.064*&0.067&0.087\\
&&&&(0.020)&(0.239)&(0.346)&(0.123)&(0.090)\\
\textit{Control variables}&&&&&&&&\\
Entry&0.113***&0.069***&0.112***&0.067**&0.239*&0.130&0.247**&0.094***\\
&(0.029)&(0.020)&(0.029)&(0.019)&(0.153)&(0.205)&(0.098)&(0.045)\\
Recession&-0.001&0.011&-0.000&0.011&-0.053***&0.069*&-0.080**&-0.009\\
&(0.029)&(0.028)&(0.029)&(0.028)&(0.030)&(0.081)&(0.056)&(0.028)\\
EMP&-0.287***&-0.254***&-0.280***&-0.241***&-0.029  &0.458&0.278&-0.471***\\
&(0.048)&(0.041)&(0.047)&(0.041)&(0.377)&(0.310)&(0.570)&(0.203)\\
GFC&-0.028&-0.051**&-0.028&-0.053**&-0.070&-0.155*&-0.001&-0.018\\
&(0.019)&(0.016)&(0.019)&(0.017)&(0.050)&(0.088)&(0.063)&(0.026)\\
\hline\
Intercept&-0.019&-0.039&-0.020&-0.036&-0.530***&0.261&-0.222**&0.440***\\
&(0.032)&(0.024)&(0.033)&(0.024)&(0.125)&(0.185)&(0.099)&(0.040)\\
Time trend&Y&Y&Y&Y& Y&Y&Y&Y\\
Industry-country FE&Y&Y&Y&Y& Y&Y&Y&Y\\
Year FE&Y&Y&Y&Y& Y&Y&Y&Y\\
R2 (within)&0.113&0.186&0.115&0.189&0.395&0.663&0.515&0.201\\
N (Groups)&5378&537&537&537&48&16&32&95\\
N (Observations)&8032&8032&8032&8032&720&240&480&1419\\
\hline\hline
\end{tabular}
}
\end{table}


\begin{thebibliography}{10}
\urlstyle{rm}
\expandafter\ifx\csname url\endcsname\relax
  \def\url#1{\texttt{#1}}\fi
\expandafter\ifx\csname urlprefix\endcsname\relax\def\urlprefix{URL }\fi
\expandafter\ifx\csname doiprefix\endcsname\relax\def\doiprefix{DOI: }\fi
\providecommand{\bibinfo}[2]{#2}
\providecommand{\eprint}[2][]{\url{#2}}

\bibitem{Solow1956}
\bibinfo{author}{Solow, R.~M.}
\newblock \bibinfo{journal}{\bibinfo{title}{{A Contribution to the Theory of
  Economic Growth}}}.
\newblock {\emph{\JournalTitle{The Quarterly Journal of Economics}}}
  \textbf{\bibinfo{volume}{70}}, \bibinfo{pages}{65},
  \doiprefix\url{10.2307/1884513} (\bibinfo{year}{1956}).

\bibitem{Knox1960}
\bibinfo{author}{Knox, A.~D.} \& \bibinfo{author}{Myrdal, G.}
\newblock \bibinfo{journal}{\bibinfo{title}{{Economic Theory and
  Under-Developed Regions.}}}
\newblock {\emph{\JournalTitle{Economica}}} \textbf{\bibinfo{volume}{27}},
  \bibinfo{pages}{280}, \doiprefix\url{10.2307/2601684} (\bibinfo{year}{1960}).

\bibitem{rodriguez2018revenge}
\bibinfo{author}{Rodr{\'{i}}guez-Pose, A.}
\newblock \bibinfo{journal}{\bibinfo{title}{{The revenge of the places that
  don't matter (and what to do about it)}}}.
\newblock {\emph{\JournalTitle{Cambridge Journal of Regions, Economy and
  Society}}} \textbf{\bibinfo{volume}{11}}, \bibinfo{pages}{189--209},
  \doiprefix\url{10.1093/cjres/rsx024} (\bibinfo{year}{2018}).

\bibitem{smith2002networks}
\bibinfo{author}{Smith, A.} \emph{et~al.}
\newblock \bibinfo{journal}{\bibinfo{title}{{Networks of value, commodities and
  regions: Reworking divisions of labour in macro-regional economies}}}.
\newblock {\emph{\JournalTitle{Progress in Human Geography}}}
  \textbf{\bibinfo{volume}{26}}, \bibinfo{pages}{41--63},
  \doiprefix\url{10.1191/0309132502ph355ra} (\bibinfo{year}{2002}).

\bibitem{Palan2010}
\bibinfo{author}{Palan, N.} \& \bibinfo{author}{Schmiedeberg, C.}
\newblock \bibinfo{journal}{\bibinfo{title}{{Structural convergence of European
  countries}}}.
\newblock {\emph{\JournalTitle{Structural Change and Economic Dynamics}}}
  \textbf{\bibinfo{volume}{21}}, \bibinfo{pages}{85--100},
  \doiprefix\url{10.1016/j.strueco.2010.01.001} (\bibinfo{year}{2010}).

\bibitem{MIDELFART2003}
\bibinfo{author}{MIDELFART, K.-H.}, \bibinfo{author}{OVERMAN, H.~G.} \&
  \bibinfo{author}{VENABLES, A.~J.}
\newblock \bibinfo{journal}{\bibinfo{title}{{Monetary Union and the Economic
  Geography of Europe}}}.
\newblock {\emph{\JournalTitle{JCMS: Journal of Common Market Studies}}}
  \textbf{\bibinfo{volume}{41}}, \bibinfo{pages}{847--868},
  \doiprefix\url{10.1111/j.1468-5965.2003.00466.x} (\bibinfo{year}{2003}).

\bibitem{Barro1991}
\bibinfo{author}{Barro, R.~J.}, \bibinfo{author}{Sala-I-Martin, X.},
  \bibinfo{author}{Blanchard, O.~J.} \& \bibinfo{author}{Hall, R.~E.}
\newblock \bibinfo{journal}{\bibinfo{title}{{Convergence Across States and
  Regions}}}.
\newblock {\emph{\JournalTitle{Brookings Papers on Economic Activity}}}
  \textbf{\bibinfo{volume}{1991}}, \bibinfo{pages}{107},
  \doiprefix\url{10.2307/2534639} (\bibinfo{year}{1991}).

\bibitem{Borsi2015}
\bibinfo{author}{Borsi, M.~T.} \& \bibinfo{author}{Metiu, N.}
\newblock \bibinfo{journal}{\bibinfo{title}{{The evolution of economic
  convergence in the European Union}}}.
\newblock {\emph{\JournalTitle{Empirical Economics}}}
  \textbf{\bibinfo{volume}{48}}, \bibinfo{pages}{657--681},
  \doiprefix\url{10.1007/s00181-014-0801-2} (\bibinfo{year}{2015}).

\bibitem{Kutan2007}
\bibinfo{author}{Kutan, A.~M.} \& \bibinfo{author}{Yigit, T.~M.}
\newblock \bibinfo{journal}{\bibinfo{title}{{European integration, productivity
  growth and real convergence}}}.
\newblock {\emph{\JournalTitle{European Economic Review}}}
  \textbf{\bibinfo{volume}{51}}, \bibinfo{pages}{1370--1395},
  \doiprefix\url{10.1016/j.euroecorev.2006.11.001} (\bibinfo{year}{2007}).

\bibitem{Kutan2009}
\bibinfo{author}{Kutan, A.~M.} \& \bibinfo{author}{Yigit, T.~M.}
\newblock \bibinfo{journal}{\bibinfo{title}{{European integration, productivity
  growth and real convergence: Evidence from the new member states}}}.
\newblock {\emph{\JournalTitle{Economic Systems}}}
  \textbf{\bibinfo{volume}{33}}, \bibinfo{pages}{127--137},
  \doiprefix\url{10.1016/j.ecosys.2009.03.002} (\bibinfo{year}{2009}).

\bibitem{Petrakos2005}
\bibinfo{author}{Petrakos, G.}, \bibinfo{author}{Rodr{\'{i}}guez-Pose, A.} \&
  \bibinfo{author}{Rovolis, A.}
\newblock \bibinfo{journal}{\bibinfo{title}{{Growth, Integration, and Regional
  Disparities in the European Union}}}.
\newblock {\emph{\JournalTitle{Environment and Planning A: Economy and Space}}}
  \textbf{\bibinfo{volume}{37}}, \bibinfo{pages}{1837--1855},
  \doiprefix\url{10.1068/a37348} (\bibinfo{year}{2005}).

\bibitem{henrekson1997growth}
\bibinfo{author}{Henrekson, M.}, \bibinfo{author}{Torstensson, J.} \&
  \bibinfo{author}{Torstensson, R.}
\newblock \bibinfo{journal}{\bibinfo{title}{{Growth effects of European
  integration}}}.
\newblock {\emph{\JournalTitle{European Economic Review}}}
  \textbf{\bibinfo{volume}{41}}, \bibinfo{pages}{1537--1557},
  \doiprefix\url{10.1016/S0014-2921(97)00063-9} (\bibinfo{year}{1997}).

\bibitem{cabral2019europe}
\bibinfo{author}{Cabral, R.} \& \bibinfo{author}{Castellanos-Sosa, F.~A.}
\newblock \bibinfo{journal}{\bibinfo{title}{{Europe's income convergence and
  the latest global financial crisis}}}.
\newblock {\emph{\JournalTitle{Research in Economics}}}
  \textbf{\bibinfo{volume}{73}}, \bibinfo{pages}{23--34},
  \doiprefix\url{10.1016/j.rie.2019.01.003} (\bibinfo{year}{2019}).

\bibitem{benczes2015european}
\bibinfo{author}{Benczes, I.} \& \bibinfo{author}{Szent-Ivanyi, B.}
\newblock \bibinfo{journal}{\bibinfo{title}{{The European Economy in 2014:
  Fragile Recovery and Convergence}}}.
\newblock {\emph{\JournalTitle{JCMS: Journal of Common Market Studies}}}
  \textbf{\bibinfo{volume}{53}}, \bibinfo{pages}{162--180},
  \doiprefix\url{10.1111/jcms.12266} (\bibinfo{year}{2015}).

\bibitem{nagy2022european}
\bibinfo{author}{Nagy, S.~G.} \& \bibinfo{author}{{\v{S}}iljak, D.}
\newblock \bibinfo{journal}{\bibinfo{title}{{Is the European Union still a
  convergence machine?}}}
\newblock {\emph{\JournalTitle{Acta Oeconomica}}}
  \textbf{\bibinfo{volume}{72}}, \bibinfo{pages}{47--63},
  \doiprefix\url{10.1556/032.2022.00003} (\bibinfo{year}{2022}).

\bibitem{crestanello2011industrial}
\bibinfo{author}{Crestanello, P.} \& \bibinfo{author}{Tattara, G.}
\newblock \bibinfo{journal}{\bibinfo{title}{{Industrial Clusters and the
  Governance of the Global Value Chain: The Romania--Veneto Network in Footwear
  and Clothing}}}.
\newblock {\emph{\JournalTitle{Regional Studies}}}
  \textbf{\bibinfo{volume}{45}}, \bibinfo{pages}{187--203},
  \doiprefix\url{10.1080/00343401003596299} (\bibinfo{year}{2011}).

\bibitem{frigant2009modular}
\bibinfo{author}{Frigant, V.} \& \bibinfo{author}{Layan, J.-B.}
\newblock \bibinfo{journal}{\bibinfo{title}{{Modular Production and the New
  Division of Labour Within Europe}}}.
\newblock {\emph{\JournalTitle{European Urban and Regional Studies}}}
  \textbf{\bibinfo{volume}{16}}, \bibinfo{pages}{11--25},
  \doiprefix\url{10.1177/0969776408098930} (\bibinfo{year}{2009}).

\bibitem{Crespo2012}
\bibinfo{author}{Crespo, N.} \& \bibinfo{author}{Sim{\~{o}}es, N.}
\newblock \bibinfo{journal}{\bibinfo{title}{{On the measurement of a
  multidimensional concept of structural similarity}}}.
\newblock {\emph{\JournalTitle{Economics Letters}}}
  \textbf{\bibinfo{volume}{116}}, \bibinfo{pages}{115--117},
  \doiprefix\url{10.1016/j.econlet.2012.01.024} (\bibinfo{year}{2012}).

\bibitem{Kallioras2010}
\bibinfo{author}{Kallioras, D.} \& \bibinfo{author}{Petrakos, G.}
\newblock \bibinfo{journal}{\bibinfo{title}{{Industrial growth, economic
  integration and structural change: evidence from the EU new member-states
  regions}}}.
\newblock {\emph{\JournalTitle{The Annals of Regional Science}}}
  \textbf{\bibinfo{volume}{45}}, \bibinfo{pages}{667--680},
  \doiprefix\url{10.1007/s00168-009-0296-5} (\bibinfo{year}{2010}).

\bibitem{Balassa1965}
\bibinfo{author}{Balassa, B.}
\newblock \bibinfo{journal}{\bibinfo{title}{{Trade Liberalisation and
  "Revealed" Comparative Advantage}}}.
\newblock {\emph{\JournalTitle{The Manchester School}}}
  \textbf{\bibinfo{volume}{33}}, \bibinfo{pages}{99--123},
  \doiprefix\url{10.1111/j.1467-9957.1965.tb00050.x} (\bibinfo{year}{1965}).

\bibitem{leamer1984sources}
\bibinfo{author}{Steuer, M.~D.} \& \bibinfo{author}{Leamer, E.~E.}
\newblock \emph{\bibinfo{title}{{Sources of International Comparative
  Advantage, Theory and Evidence.}}}, vol. \bibinfo{volume}{54 (214)}
  (\bibinfo{publisher}{MIT press Cambridge, MA}, \bibinfo{year}{1987}).

\bibitem{krugman1996integration}
\bibinfo{author}{Krugman, P.} \& \bibinfo{author}{Venables, A.~J.}
\newblock \bibinfo{journal}{\bibinfo{title}{{Integration, specialization, and
  adjustment}}}.
\newblock {\emph{\JournalTitle{European Economic Review}}}
  \textbf{\bibinfo{volume}{40}}, \bibinfo{pages}{959--967},
  \doiprefix\url{10.1016/0014-2921(95)00104-2} (\bibinfo{year}{1996}).

\bibitem{romer1987growth}
\bibinfo{author}{Romer, P.}
\newblock \bibinfo{journal}{\bibinfo{title}{{Growth Based on Increasing Returns
  Due to Specialization}}}.
\newblock {\emph{\JournalTitle{The American Economic Review}}}
  \textbf{\bibinfo{volume}{77}}, \bibinfo{pages}{56--62},
  \doiprefix\url{10.2307/1805429} (\bibinfo{year}{1987}).

\bibitem{marshall2009principles}
\bibinfo{author}{Marshall, A.}
\newblock \emph{\bibinfo{title}{{Principles of Economics (Eighth Edition)}}}
  (\bibinfo{publisher}{Cosimo, Inc.}, \bibinfo{year}{1890}).

\bibitem{archibugi1994aggregate}
\bibinfo{author}{Archibugi, D.} \& \bibinfo{author}{Pianta, M.}
\newblock \bibinfo{journal}{\bibinfo{title}{{Aggregate convergence and sectoral
  specialization in innovation}}}.
\newblock {\emph{\JournalTitle{Journal of Evolutionary Economics}}}
  \textbf{\bibinfo{volume}{4}}, \bibinfo{pages}{17--33},
  \doiprefix\url{10.1007/BF01200835} (\bibinfo{year}{1994}).

\bibitem{imbs2003stages}
\bibinfo{author}{Imbs, J.} \& \bibinfo{author}{Wacziarg, R.}
\newblock \bibinfo{journal}{\bibinfo{title}{{Stages of Diversification}}}.
\newblock {\emph{\JournalTitle{American Economic Review}}}
  \textbf{\bibinfo{volume}{93}}, \bibinfo{pages}{63--86},
  \doiprefix\url{10.1257/000282803321455160} (\bibinfo{year}{2003}).

\bibitem{Hidalgo:2007aa}
\bibinfo{author}{Hidalgo, C.~A.}, \bibinfo{author}{Klinger, B.},
  \bibinfo{author}{Barab{\'a}si, A.-L.} \& \bibinfo{author}{Hausmann, R.}
\newblock \bibinfo{journal}{\bibinfo{title}{{The Product Space Conditions the
  Development of Nations}}}.
\newblock {\emph{\JournalTitle{Science}}} \textbf{\bibinfo{volume}{317}},
  \bibinfo{pages}{482--487}, \doiprefix\url{10.1126/science.1144581}
  (\bibinfo{year}{2007}).

\bibitem{Hidalgo:2009aa}
\bibinfo{author}{Hidalgo, C.~A.} \& \bibinfo{author}{Hausmann, R.}
\newblock \bibinfo{journal}{\bibinfo{title}{{The building blocks of economic
  complexity}}}.
\newblock {\emph{\JournalTitle{Proceedings of the National Academy of
  Sciences}}} \textbf{\bibinfo{volume}{106}}, \bibinfo{pages}{10570--10575},
  \doiprefix\url{10.1073/pnas.0900943106} (\bibinfo{year}{2009}).

\bibitem{fontagne2008specialization}
\bibinfo{author}{Fontagn, L.}, \bibinfo{author}{Gaulier, G.} \&
  \bibinfo{author}{Zignago, S.}
\newblock \bibinfo{journal}{\bibinfo{title}{{Specialization across Varieties
  and North--South Competition}}}.
\newblock {\emph{\JournalTitle{Economic Policy}}}
  \textbf{\bibinfo{volume}{23}}, \bibinfo{pages}{53--93},
  \doiprefix\url{10.1002/9781444306699.ch2} (\bibinfo{year}{2009}).

\bibitem{levchenko2012comparative}
\bibinfo{author}{Levchenko, A.~A.} \& \bibinfo{author}{Zhang, J.}
\newblock \bibinfo{journal}{\bibinfo{title}{{Comparative advantage and the
  welfare impact of European integration}}}.
\newblock {\emph{\JournalTitle{Economic Policy}}}
  \textbf{\bibinfo{volume}{27}}, \bibinfo{pages}{567--602},
  \doiprefix\url{10.1111/j.1468-0327.2012.00294.x} (\bibinfo{year}{2012}).

\bibitem{pop00008}
\bibinfo{author}{{Di Clemente}, R.}, \bibinfo{author}{Chiarotti, G.~L.},
  \bibinfo{author}{Cristelli, M.}, \bibinfo{author}{Tacchella, A.} \&
  \bibinfo{author}{Pietronero, L.}
\newblock \bibinfo{journal}{\bibinfo{title}{{Diversification versus
  Specialization in Complex Ecosystems}}}.
\newblock {\emph{\JournalTitle{PLoS ONE}}} \textbf{\bibinfo{volume}{9}},
  \bibinfo{pages}{e112525}, \doiprefix\url{10.1371/journal.pone.0112525}
  (\bibinfo{year}{2014}).

\bibitem{saracco2015}
\bibinfo{author}{Saracco, F.}, \bibinfo{author}{{Di Clemente}, R.},
  \bibinfo{author}{Gabrielli, A.} \& \bibinfo{author}{Pietronero, L.}
\newblock \bibinfo{journal}{\bibinfo{title}{{From innovation to
  diversification: A simple competitive model}}}.
\newblock {\emph{\JournalTitle{PLoS ONE}}} \textbf{\bibinfo{volume}{10}},
  \bibinfo{pages}{e0140420}, \doiprefix\url{10.1371/journal.pone.0140420}
  (\bibinfo{year}{2015}).
\newblock \eprint{1508.03571}.

\bibitem{Zweig2011}
\bibinfo{author}{Zweig, K.~A.} \& \bibinfo{author}{Kaufmann, M.}
\newblock \bibinfo{journal}{\bibinfo{title}{{A systematic approach to the
  one-mode projection of bipartite graphs}}}.
\newblock {\emph{\JournalTitle{Social Network Analysis and Mining}}}
  \textbf{\bibinfo{volume}{1}}, \bibinfo{pages}{187--218},
  \doiprefix\url{10.1007/s13278-011-0021-0} (\bibinfo{year}{2011}).

\bibitem{Milo2002}
\bibinfo{author}{Milo, R.}
\newblock \bibinfo{journal}{\bibinfo{title}{{Network Motifs: Simple Building
  Blocks of Complex Networks}}}.
\newblock {\emph{\JournalTitle{Science}}} \textbf{\bibinfo{volume}{298}},
  \bibinfo{pages}{824--827}, \doiprefix\url{10.1126/science.298.5594.824}
  (\bibinfo{year}{2002}).

\bibitem{Simmons2019}
\bibinfo{author}{Simmons, B.~I.} \emph{et~al.}
\newblock \bibinfo{journal}{\bibinfo{title}{{bmotif: A package for motif
  analyses of bipartite networks}}}.
\newblock {\emph{\JournalTitle{Methods in Ecology and Evolution}}}
  \textbf{\bibinfo{volume}{10}}, \bibinfo{pages}{695--701},
  \doiprefix\url{10.1111/2041-210X.13149} (\bibinfo{year}{2019}).

\bibitem{Dormann2009}
\bibinfo{author}{Dormann, C.~F.}, \bibinfo{author}{Frund, J.},
  \bibinfo{author}{Bluthgen, N.} \& \bibinfo{author}{Gruber, B.}
\newblock \bibinfo{journal}{\bibinfo{title}{{Indices, Graphs and Null Models:
  Analyzing Bipartite Ecological Networks}}}.
\newblock {\emph{\JournalTitle{The Open Ecology Journal}}}
  \textbf{\bibinfo{volume}{2}}, \bibinfo{pages}{7--24},
  \doiprefix\url{10.2174/1874213000902010007} (\bibinfo{year}{2009}).

\bibitem{Saracco2015a}
\bibinfo{author}{Saracco, F.}, \bibinfo{author}{{Di Clemente}, R.},
  \bibinfo{author}{Gabrielli, A.} \& \bibinfo{author}{Squartini, T.}
\newblock \bibinfo{journal}{\bibinfo{title}{{Randomizing bipartite networks:
  the case of the World Trade Web}}}.
\newblock {\emph{\JournalTitle{Scientific Reports}}}
  \textbf{\bibinfo{volume}{5}}, \bibinfo{pages}{10595},
  \doiprefix\url{10.1038/srep10595} (\bibinfo{year}{2015}).

\bibitem{Gualdi2016}
\bibinfo{author}{Gualdi, S.}, \bibinfo{author}{Cimini, G.},
  \bibinfo{author}{Primicerio, K.}, \bibinfo{author}{{Di Clemente}, R.} \&
  \bibinfo{author}{Challet, D.}
\newblock \bibinfo{journal}{\bibinfo{title}{{Statistically validated network of
  portfolio overlaps and systemic risk}}}.
\newblock {\emph{\JournalTitle{Scientific Reports}}}
  \textbf{\bibinfo{volume}{6}}, \bibinfo{pages}{39467},
  \doiprefix\url{10.1038/srep39467} (\bibinfo{year}{2016}).
\newblock \eprint{1603.05914}.

\bibitem{Cimini2014}
\bibinfo{author}{Cimini, G.}, \bibinfo{author}{Gabrielli, A.} \&
  \bibinfo{author}{{Sylos Labini}, F.}
\newblock \bibinfo{journal}{\bibinfo{title}{{The Scientific Competitiveness of
  Nations}}}.
\newblock {\emph{\JournalTitle{PLoS ONE}}} \textbf{\bibinfo{volume}{9}},
  \bibinfo{pages}{e113470}, \doiprefix\url{10.1371/journal.pone.0113470}
  (\bibinfo{year}{2014}).
\newblock \eprint{1409.5698}.

\bibitem{Saracco2016}
\bibinfo{author}{Saracco, F.}, \bibinfo{author}{{Di Clemente}, R.},
  \bibinfo{author}{Gabrielli, A.} \& \bibinfo{author}{Squartini, T.}
\newblock \bibinfo{journal}{\bibinfo{title}{{Detecting early signs of the
  2007-2008 crisis in the world trade}}}.
\newblock {\emph{\JournalTitle{Scientific Reports}}}
  \textbf{\bibinfo{volume}{6}}, \bibinfo{pages}{30286},
  \doiprefix\url{10.1038/srep30286} (\bibinfo{year}{2016}).

\bibitem{Clemente2020}
\bibinfo{author}{{Di Clemente}, R.}, \bibinfo{author}{Strano, E.} \&
  \bibinfo{author}{Batty, M.}
\newblock \bibinfo{journal}{\bibinfo{title}{{Urbanization and economic
  complexity}}}.
\newblock {\emph{\JournalTitle{Scientific Reports}}}
  \textbf{\bibinfo{volume}{11}}, \bibinfo{pages}{3952},
  \doiprefix\url{10.1038/s41598-021-83238-5} (\bibinfo{year}{2021}).
\newblock \eprint{2009.11966}.

\bibitem{Metcalfe2006}
\bibinfo{author}{Metcalfe, J.~S.}, \bibinfo{author}{Foster, J.} \&
  \bibinfo{author}{Ramlogan, R.}
\newblock \bibinfo{journal}{\bibinfo{title}{{Adaptive economic growth}}}.
\newblock {\emph{\JournalTitle{Cambridge Journal of Economics}}}
  \textbf{\bibinfo{volume}{30}}, \bibinfo{pages}{7--32},
  \doiprefix\url{10.1093/cje/bei055} (\bibinfo{year}{2006}).

\bibitem{tacchella2012new}
\bibinfo{author}{Tacchella, A.}, \bibinfo{author}{Cristelli, M.},
  \bibinfo{author}{Caldarelli, G.}, \bibinfo{author}{Gabrielli, A.} \&
  \bibinfo{author}{Pietronero, L.}
\newblock \bibinfo{journal}{\bibinfo{title}{{A New Metrics for Countries'
  Fitness and Products' Complexity}}}.
\newblock {\emph{\JournalTitle{Scientific Reports}}}
  \textbf{\bibinfo{volume}{2}}, \bibinfo{pages}{723},
  \doiprefix\url{10.1038/srep00723} (\bibinfo{year}{2012}).

\bibitem{Saracco2017}
\bibinfo{author}{Saracco, F.} \emph{et~al.}
\newblock \bibinfo{journal}{\bibinfo{title}{{Inferring monopartite projections
  of bipartite networks: An entropy-based approach}}}.
\newblock {\emph{\JournalTitle{New Journal of Physics}}}
  \textbf{\bibinfo{volume}{19}}, \bibinfo{pages}{053022},
  \doiprefix\url{10.1088/1367-2630/aa6b38} (\bibinfo{year}{2017}).
\newblock \eprint{1607.02481}.

\bibitem{VanLidthdeJeude2019}
\bibinfo{author}{{van Lidth de Jeude}, J.}, \bibinfo{author}{{Di Clemente},
  R.}, \bibinfo{author}{Caldarelli, G.}, \bibinfo{author}{Saracco, F.} \&
  \bibinfo{author}{Squartini, T.}
\newblock \bibinfo{journal}{\bibinfo{title}{{Reconstructing Mesoscale Network
  Structures}}}.
\newblock {\emph{\JournalTitle{Complexity}}} \textbf{\bibinfo{volume}{2019}},
  \bibinfo{pages}{1--13}, \doiprefix\url{10.1155/2019/5120581}
  (\bibinfo{year}{2019}).

\bibitem{kolev2012impact}
\bibinfo{author}{Kolev, A.}
\newblock \bibinfo{journal}{\bibinfo{title}{{The impact of the recession in
  2008-2009 on EU regional convergence}}}.
\newblock {\emph{\JournalTitle{ECON Department, European Investment Bank,
  Economic Studies Division SG/ECON/ES/2012-522, http://www. eib.
  org/infocentre/publications/all/econ-note-2012-regional-convergence.
  htm,(Eri{\c{s}}im, 02.12. 2012)}}} \bibinfo{pages}{12}
  (\bibinfo{year}{2012}).

\bibitem{HRISTOV2012569}
\bibinfo{author}{Hristov, N.}, \bibinfo{author}{H{\"{u}}lsewig, O.} \&
  \bibinfo{author}{Wollmersh{\"{a}}user, T.}
\newblock \bibinfo{journal}{\bibinfo{title}{{Loan supply shocks during the
  financial crisis: Evidence for the Euro area}}}.
\newblock {\emph{\JournalTitle{Journal of International Money and Finance}}}
  \textbf{\bibinfo{volume}{31}}, \bibinfo{pages}{569--592},
  \doiprefix\url{10.1016/j.jimonfin.2011.10.007} (\bibinfo{year}{2012}).

\bibitem{szekely2009economic}
\bibinfo{author}{Szekely, I.} \& \bibinfo{author}{van~den Noord, P.~J.}
\newblock \bibinfo{journal}{\bibinfo{title}{{Economic crisis in Europe: cause,
  consequences, and responses}}}.
\newblock {\emph{\JournalTitle{A report by the European Commission. European
  Commission, Directorate-General for Economic and Financial Affairs}}}
  (\bibinfo{year}{2009}).

\bibitem{pavlinek2002transformation}
\bibinfo{author}{Pavl{\'{i}}nek, P.}
\newblock \bibinfo{journal}{\bibinfo{title}{{Transformation of the Central and
  East European Passenger Car Industry: Selective Peripheral Integration
  through Foreign Direct Investment}}}.
\newblock {\emph{\JournalTitle{Environment and Planning A: Economy and Space}}}
  \textbf{\bibinfo{volume}{34}}, \bibinfo{pages}{1685--1709},
  \doiprefix\url{10.1068/a34263} (\bibinfo{year}{2002}).

\bibitem{cutrini2022eu}
\bibinfo{author}{Cutrini, E.}, \bibinfo{author}{Gardiner, B.} \&
  \bibinfo{author}{Martin, R.}
\newblock \bibinfo{journal}{\bibinfo{title}{{EU integration and the geographies
  of economic activity: 1985--2019}}}.
\newblock {\emph{\JournalTitle{Environment and Planning A: Economy and Space}}}
  \bibinfo{pages}{0308518X2211270}, \doiprefix\url{10.1177/0308518X221127028}
  (\bibinfo{year}{2022}).

\bibitem{du2010inter}
\bibinfo{author}{{Du Caju}, P.}, \bibinfo{author}{K{\'{a}}tay, G.},
  \bibinfo{author}{Lamo, A.}, \bibinfo{author}{Nicolitsas, D.} \&
  \bibinfo{author}{Poelhekke, S.}
\newblock \bibinfo{journal}{\bibinfo{title}{{Inter-Industry Wage Differentials
  In EU Countries: What Do Cross-Country Time Varying Data Add to the
  Picture?}}}
\newblock {\emph{\JournalTitle{Journal of the European Economic Association}}}
  \textbf{\bibinfo{volume}{8}}, \bibinfo{pages}{478--486},
  \doiprefix\url{10.1162/jeea.2010.8.2-3.478} (\bibinfo{year}{2010}).

\bibitem{grossman2012task}
\bibinfo{author}{Grossman, G.~M.} \& \bibinfo{author}{Rossi-Hansberg, E.}
\newblock \bibinfo{journal}{\bibinfo{title}{{Task Trade Between Similar
  Countries}}}.
\newblock {\emph{\JournalTitle{Econometrica}}} \textbf{\bibinfo{volume}{80}},
  \bibinfo{pages}{593--629}, \doiprefix\url{10.3982/ECTA8700}
  (\bibinfo{year}{2012}).

\bibitem{neary2009putting}
\bibinfo{author}{Neary, J.~P.}
\newblock \bibinfo{journal}{\bibinfo{title}{{Putting the ``New'' into New Trade
  Theory: Paul Krugman's Nobel Memorial Prize in Economics}}}.
\newblock {\emph{\JournalTitle{Scandinavian Journal of Economics}}}
  \textbf{\bibinfo{volume}{111}}, \bibinfo{pages}{217--250},
  \doiprefix\url{10.1111/j.1467-9442.2009.01562.x} (\bibinfo{year}{2009}).

\bibitem{helpman1987market}
\bibinfo{author}{Helpman, E.} \& \bibinfo{author}{Krugman, P.}
\newblock \emph{\bibinfo{title}{{Market structure and foreign trade: Increasing
  returns, imperfect competition, and the international economy}}}
  (\bibinfo{publisher}{MIT press}, \bibinfo{year}{1987}).

\bibitem{McCann2015}
\bibinfo{author}{McCann, P.} \& \bibinfo{author}{Ortega-Argil{\'{e}}s, R.}
\newblock \bibinfo{journal}{\bibinfo{title}{{Smart Specialization, Regional
  Growth and Applications to European Union Cohesion Policy}}}.
\newblock {\emph{\JournalTitle{Regional Studies}}}
  \textbf{\bibinfo{volume}{49}}, \bibinfo{pages}{1291--1302},
  \doiprefix\url{10.1080/00343404.2013.799769} (\bibinfo{year}{2015}).

\bibitem{Thissen2013}
\bibinfo{author}{Thissen, M.}, \bibinfo{author}{van Oort, F.},
  \bibinfo{author}{Diodato, D.} \& \bibinfo{author}{Ruijs, A.}
\newblock \emph{\bibinfo{title}{{Regional Competitiveness and Smart
  Specialization in Europe}}} (\bibinfo{publisher}{Edward Elgar Publishing},
  \bibinfo{year}{2013}).

\bibitem{Capello2016}
\bibinfo{author}{Capello, R.} \& \bibinfo{author}{Kroll, H.}
\newblock \bibinfo{journal}{\bibinfo{title}{{From theory to practice in smart
  specialization strategy: emerging limits and possible future trajectories}}}.
\newblock {\emph{\JournalTitle{European Planning Studies}}}
  \textbf{\bibinfo{volume}{24}}, \bibinfo{pages}{1393--1406},
  \doiprefix\url{10.1080/09654313.2016.1156058} (\bibinfo{year}{2016}).

\bibitem{Cimini2019}
\bibinfo{author}{Cimini, G.} \emph{et~al.}
\newblock \bibinfo{journal}{\bibinfo{title}{{The statistical physics of
  real-world networks}}}.
\newblock {\emph{\JournalTitle{Nature Reviews Physics}}}
  \textbf{\bibinfo{volume}{1}}, \bibinfo{pages}{58--71},
  \doiprefix\url{10.1038/s42254-018-0002-6} (\bibinfo{year}{2019}).
\newblock \eprint{1810.05095}.

\bibitem{Holland1981}
\bibinfo{author}{Holland, P.~W.} \& \bibinfo{author}{Leinhardt, S.}
\newblock \bibinfo{journal}{\bibinfo{title}{{An Exponential Family of
  Probability Distributions for Directed Graphs}}}.
\newblock {\emph{\JournalTitle{Journal of the American Statistical
  Association}}} \textbf{\bibinfo{volume}{76}}, \bibinfo{pages}{33--50},
  \doiprefix\url{10.1080/01621459.1981.10477598} (\bibinfo{year}{1981}).

\bibitem{straka2017}
\bibinfo{author}{Straka, M.~J.}, \bibinfo{author}{Caldarelli, G.} \&
  \bibinfo{author}{Saracco, F.}
\newblock \bibinfo{journal}{\bibinfo{title}{{Grand canonical validation of the
  bipartite international trade network}}}.
\newblock {\emph{\JournalTitle{Physical Review E}}}
  \textbf{\bibinfo{volume}{96}}, \bibinfo{pages}{022306},
  \doiprefix\url{10.1103/PhysRevE.96.022306} (\bibinfo{year}{2017}).

\bibitem{Payrato-Borras2019}
\bibinfo{author}{Payrat{\'{o}}-Borr{\`{a}}s, C.},
  \bibinfo{author}{Hern{\'{a}}ndez, L.} \& \bibinfo{author}{Moreno, Y.}
\newblock \bibinfo{journal}{\bibinfo{title}{{Breaking the Spell of Nestedness:
  The Entropic Origin of Nestedness in Mutualistic Systems}}}.
\newblock {\emph{\JournalTitle{Physical Review X}}}
  \textbf{\bibinfo{volume}{9}}, \bibinfo{pages}{031024},
  \doiprefix\url{10.1103/PhysRevX.9.031024} (\bibinfo{year}{2019}).

\end{thebibliography}
\end{document}